\documentclass[preprint,12pt]{mycls}

\usepackage{amsmath}
\usepackage{amssymb}
\usepackage{color}
\usepackage{simplewick}

\begin{document}

\begin{frontmatter}

\title{Two-dimensional conformal field theories with matrix-valued level} 

\author{Ali Nassar}
 \ead{anassar@zewailcity.edu.eg}

\address{Department of Physics, University of Science and Technology, Zewail City of Science and Technology,
 12588 Giza, Egypt}

\begin{abstract}
We introduce a new class of two dimensional conformal field theories by extending  Wess-Zumino-Witten (WZW) models to chiral algebras with matrix-valued levels. The new CFTs are based on holomorphic currents with an operator product expansion characterized by a positive integer-valued matrix $K_{AB}$.  We use the Sugawara construction to compute the energy-momentum tensor, the central charge, and the spectrum of conformal dimensions of the CFTs based on this algebra. We also construct a set of genus-$1$ characters and show that they fulfil a representation of the modular group $\text{SL}(2,\mathbb{Z})$ up to a modular anomaly.

\end{abstract}

\end{frontmatter}

\section{Introduction}

The study of 2D conformal field theories (CFTs) is motivated by their importance in string theory and critical phenomena (see \cite{Belavin:1984vu,bigbook}). They describe the world-sheet dynamics of perturbative string theory and correspond to string backgrounds which solves the classical equations of motion \cite{Polchinski:1998rq,Polchinski:1998rq2}. In many statistical mechanical systems, CFTs describe renormalization group fixed  points when the system exhibits scale invariance and long-range correlations. One of the major unsolved problems in this field is the construction and  classification  of all 2D CFTs (see \cite{Gannon:1999bi,Wendland:2000ye} for a review and references therein).
An important structure in 2D CFT is the chiral algebra which gives rise to infinite-dimensional chiral symmetries. This chiral algebra
could be the Virasoro algebra or one of its extensions. The Virasoro algebra $\text{Vir}\oplus \overline{\text{Vir}}$ is the chiral algebra which underlies any 2D CFT and is generated by the energy-momentum tensor $T(z), \overline{T}(\overline{z})$ which is a spin-2 field. The chiral algebra can get extended by the inclusion of integer- or half-integer-spin holomorphic fields\footnote{Half-integer spin fields generate superconformal symmetries.}  \cite{Bouwknegt:1992wg}. The possible extension of the Virasoro algebra have been classified by Zamolodchikov \cite{Zamolodchikov:1985wn}.  In this paper, we consider the extension of the Virasoro algebra by spin-1 holomorphic currents $J_A^a(z)$ with an operator product expansion (OPE) characterized by a matrix-valued level $K_{AB}$. The zero modes of these currents generate the horizontal subalgebra $g^{\oplus_N}=g\oplus\cdots\oplus g$ where  $g$ is a compact simple Lie algebra. We will denote this algebra by $\hat{g}_{N,K}$.  For a diagonal matrix-level $k_A \delta_{AB}$ the algebra $\hat{g}_{N,K}$ reduces to a  Wess-Zumino-Witten (WZW) model based on $g\oplus\cdots\oplus g$. The algebra $\hat{g}_{N,K}$ is not the central extension of the loop algebra of $g^{\oplus_N}$ due to a cohomological obstruction.
We will use the Sugawara construction to construct an energy-momentum tensor as a bilinear in the currents $J_A^a(z)$. This leads to  2D CFTs with new values for the central and charge conformal dimensions. A set of specialized genus-1 characters will be constructed and we will show that they transform under the modular group $\text{SL}(2,\mathbb{Z})$ up to a modular anomaly proportional to the above mentioned cohomological obstruction.

 This paper is organized as follows: In Section (\ref{sec2}), we introduce the matrix-valued algebra $\hat{g}_{N,K}$ and compute its central charge and conformal dimensions. In Section (\ref{sec3}), we write down the genus-1 characters and show that they transform under the modular group $\text{SL}(2,\mathbb{Z})$ up to a modular anomaly. In Section (\ref{sec4}), we apply our results to the algebra $\widehat{su(2)}_{N,K}$.

\section{The matrix-level algebra $\hat{g}_{N,K}$}\label{sec2}

One source of chiral algebras is the affine Kac-Moody current algebras based on the central extension of some Lie algebra \cite{Kac:1990gs}, e.g., the level $k$ WZW models based on $\hat{g}_k$ \cite{Witten:1983ar}. For a compact simply-connected Lie group $G$ with a simple Lie algebra $g$, the Kac-Moody currents $J^a(z)$ of $\hat{g}_k$ have an operator product expansion (OPE)
\begin{equation}
J^a(z) J^b(w)\sim \frac{if_{abc} J^c(w) }{z-w}+\frac{k\delta^{ab}}{(z-w)^2}.
\end{equation}
This OPE gives rise to following modes algebra
\begin{equation}
[J^a_m,J^b_n]=if_{abc} J^c_{m+n}+k\Omega(J^a_m,J^b_n), \quad \Omega(J^a_m,J^b_n)=m\delta^{ab} \delta_{m+n,0},
\end{equation}
 where $k\in\mathbb{Z}_{+}$.

 The central extension term  $\Omega(J^a_m,J^b_n)$  defines a $2$-cocycle on the loop group of $g$, i.e., it satisfies
 \begin{equation}
 \begin{gathered}
 \Omega(J^a_m,J^b_n)=-\Omega(J^b_n, J^a_m)\\[4pt]
  \Omega([J^a_m,[J^b_n,J^c_r]])+\Omega([J^c_r,[J^a_m, J^b_n]])+\Omega([J^b_n,[J^c_r,J^a_m]])=0.
 \end{gathered}
 \end{equation}

The energy-momentum tensor is constructed by the Sugawara construction
\begin{equation}
T(z)=\gamma J^a(z) J^a(z),
\end{equation}
where $\gamma$ is fixed by the requirement that $J^a(z)$ has conformal dimension equal to one. This leads to
\begin{equation}
\gamma=\frac{1}{2(k+h)},
\end{equation}
where $h$ is the dual Coxeter number of $g$. Once $\gamma$ is fixed, one can use the $TT$ OPE to compute the central charge of the theory
\begin{equation}
T(z)T(w)\sim \frac{c/2}{(z-w)^4}+\frac{2T(w)}{(z-w)^2}+\frac{\partial T(w)}{(z-w)}
\end{equation}
which gives
\begin{equation}
c(k)=\frac{k\; \text{dim}\; g}{k+h}.
\end{equation}

The Sugawara construction can be extended to the case of semisimple Lie algebras which are direct sums $g^{\oplus_N}=g_1\oplus\cdots\oplus g_N$ of simple algebras $g_A$. The total energy-momentum tensor will be the sum of the Sugawara energy-momentum tensors for each simple factor. The fact that the individual energy-momentum tensors commute means that the total central charge will be the sum of the individual central charges
\begin{equation}
C=\sum_{A=1}^N c(k_A)=\sum_{A=1}^N \frac{k_A\;  \text{dim}\; g_A}{k_A+h_A},
\end{equation}
where $k_A \in \mathbb{Z}_{+}$, $\text{dim}\; g_A$, and $h_A$ are the level, the dimension, and the dual Coxeter number of $\hat{g}^A_{k_A}$, respectively.

We propose a matrix-level Kac-Moody algebra based on $g^{\oplus_N}$ which we denote by $\hat{g}_{N,K}$.
We take $K_{AB}$ to be the intersection form of a positive integral\footnote{See the discussion after (\ref{unitarity2})} lattice $\Gamma_K$, i.e., $K_{AB} \in \mathbb{Z}_+$. The inverse $K^{AB}$ is the intersection form of the dual lattice $\Gamma_K^*$.

We will denote the currents by $J_A^{a}$, where $a$ is the Lie algebra index $a=1,\cdots,\dim(g)$ and $A=1,\cdots,N=\text{rank}(K)$. The Lie algebra of  $g^{\oplus N}$ takes the form
\begin{equation}
\big[J_A^{a},J_B^{b}\big]=if_{abc} J_B^{c}\delta_{AB}.
\end{equation}
We assume the following OPE of the currents
\begin{equation}\label{matrix-level-alg}
J_A^{a}(z)J_B^{b}(w)\sim  \frac{if_{abc} J_B^{c}(w)\delta_{AB}}{z-w}+\frac{\delta^{ab} K_{AB}}{(z-w)^2}.
\end{equation}

An important consistency check  which we will use in our subsequent calculations is the concept of \textit{factorization}, i.e., when $K_{AB}=k_A \delta_{AB}$ (diagonal matrix level) all our formulas should boil down to the familiar algebra
\begin{equation}
\hat{g}_{N,\vec{k}}=\hat{g}_{k_1}\oplus\cdots\oplus \hat{g}_{k_N}.
\end{equation}



We use the Sugawara construction to find the energy-momentum tensor
\begin{equation}
T(z)=\sum_{A,B=1}^N G^{AB} (J_A^a J_B^a)(z),
\end{equation}
where the summation convention is used on the Lie algebra index $a$. Here $G^{AB}$ is such that the currents $J_M^a(z)$ have conformal dimension one
\begin{equation}
T(w) J_M^a (z)\sim \frac{J_M^a (z)}{(w-z)^2}+\frac{\partial J_M^a (z)}{(w-z)}.
\end{equation}
To compute this OPE we use
\begin{equation}
\begin{split}
&T(w)J_M^a(z)\\
&\qquad = \sum_{A,B=1}^N G^{AB} (J_A^b J_B^b)(w) J_M^a (z) \\[6pt]
& \qquad =\frac{1}{2\pi i} \oint_w \frac{dx}{x-w} \sum_{A,B=1}^N \bigg[G^{AB}\contraction{}{J_M^a(z)}{}{J_A^b(x)} J_M^a(z) J_A^b(x) J_B^b(w) + G^{AB} \contraction{J_M^a(z)}{J_A^b(x)}{}{J_B^b(w)} J_A^b(x) J_M^a (z)J_B^b(w)\bigg]\\[6pt]
& \qquad =\frac{1}{2\pi i} \oint_w \frac{dx}{x-w} \sum_{A,B=1}^N  \Bigg\{G^{AB} \bigg[\frac{i f_{abc} \delta_{MA} J_A^c (x) }{(z-x)} +\frac{K_{MA}\delta_{ab} }{(z-x)^2}\bigg] J_B^b(w)\\[6pt]
& \qquad \quad +G^{AB} J_A^b(x) \bigg[\frac{i f_{abc} \delta_{MB} J_B^c (w) }{(z-w)} +\frac{K_{MB}\delta_{ab} }{(z-w)^2}\bigg] \Bigg\}\\[6pt]
& \qquad =\frac{1}{2\pi i} \oint_w \frac{dx}{x-w} \sum_{A,B=1}^N  \Bigg\{ \frac{- f_{abc}  f_{cbd} G^{AB} \delta_{MA}  \delta_{AB} J_B^d (w) }{(z-x) (x-w)} + \frac{i f_{abc} G^{AB} \delta_{MA} (J_A^c J_B^b)(w)}{(z-x)}\\[6pt]
& \qquad \quad   +\frac{G^{AB} K_{MA} J_B^a(w)}{(z-x)^2}+\frac{G^{AB} K_{MB} J_A^a(x)}{(z-w)^2} +\frac{if_{abc}G^{AB}\delta_{MB} (J_A^b J_B^c)(w) }{(z-w)}  \Bigg\}
\end{split}
\end{equation}
Using
\begin{equation}
f_{abc}  f_{cbd} =-2h \delta_{ad},\qquad \sum_B G^{MB}f_{abc} \big[(J_M^c J_B^b)+(J_M^b J_B^c)  \big]=0
\end{equation}
Then
\begin{equation}\label{Gequation}
\begin{split}
&\sum_{A,B=1}^N G^{AB} (J_A^b J_B^b)(w) J_M^a (z) \\[6pt]
& \qquad =\frac{1}{2\pi i} \oint_w \frac{dx}{x-w} \sum_{A,B=1}^N \bigg[\frac{G^{AB} K_{MA} J_B^a(x)}{(z-x)^2}  +  \frac{2h G^{MM} J_M^a(w)}{(z-x)(x-w)} +\frac{ G^{AB} K_{MB} J_A^a(w)}{(z-w)^2}\bigg]\\[6pt]
& \qquad =\sum_{A,B=1}^N \frac{G^{AB} K_{MA} J_B^a(w)  +  2h G^{MM} J_M^a(w) + G^{AB} K_{MB} J_A^a(w)}{(z-w)^2}.
\end{split}
\end{equation}
We look for a  $G^{AB}$ of the form
\begin{equation}
G^{AB}=\gamma^A K ^{AB},
\end{equation}
where $K^{AB}=K_{AB}^{-1}$ and  $\gamma^A$ will be chosen in such a way to force $J_M^a$ to have conformal dimension equal to one. Using this form of $G^{AB}$ in (\ref{Gequation}) we get
\begin{equation}
\gamma^A=\frac{1}{2(1+hK^{AA})}
\end{equation}

\begin{equation}\label{Gmatrix}
G^{AB}=\frac{K^{AB}}{2(1+hK^{AA})}.
\end{equation}
 The energy-momentum tensor now reads
\begin{equation}
T(z)=\sum_{A,B=1}^N\frac{K^{AB}}{2(1+hK^{AA})} (J_A^a J_B^a)(z).
\end{equation}
The $TJ$ OPE now reads
\begin{equation}
T(z) J_A^a(w) \sim \frac{J_A^a(w)}{(z-w)^2}+\frac{\partial J_A^a(w)}{(z-w)}.
\end{equation}

The OPE of $T(z)$ with itself can be similarly computed


\begin{equation}\label{TTope}
\begin{split}
& T(z)T(w)\\[6pt] &
\quad = \sum_{A,B=1}^N G^{AB} \frac{1}{2\pi i} \oint_w \frac{dx}{x-w} \bigg[\contraction{}{T(z)}{} {J_A^b(x)} T(z)J_A^a(x)J_B^a(w)
+ \contraction{}{J_A^a(x)} {T(z)} {J_A^b(x)} J_A^a(x) T(z) J_B^a(w)\bigg]\\[6pt]
&\quad  = \sum_{A,B=1}^N G^{AB} \frac{1}{2\pi i} \oint_w \frac{dx}{x-w}\bigg\{\bigg[\frac{J_A^a(x)J_B^a(w)}{(z-x)^2} +\frac{\partial J_A^a(x)J_B^a(w)}{(z-x)}\bigg]\\[6pt] &\qquad
+ \bigg[\frac{J_A^a(x)J_B^a(w)}{(z-w)^2} +\frac{ J_A^a(x) \partial J_B^a(w)}{(z-w)}\bigg]\bigg\}\\[6pt]
&\quad = \sum_{A,B=1}^N G^{AB} \frac{1}{2\pi i} \oint_w \frac{dx}{x-w}\bigg\{\bigg[\frac{\dim g\; K_{AB}}{(z-x)^2 (x-w)^2}+\frac{(J_A^a J_B^a)(w)}{(z-x)^2}\\[4pt]&\qquad -\frac{2 \dim g\; K_{AB}}{(z-x)(x-w)^3}+ \frac{(\partial J_A^a J_B^a)(w)}{(z-x)}\bigg]+ \bigg[\frac{(J_A^a J_B^a)(w)}{(z-w)^2} + \frac{( J_A^a \partial J_B^a)(w)}{(z-w)} \bigg]\bigg\},
\end{split}
\end{equation}
where we used
\begin{equation}
\delta^{aa}=\dim g, \qquad J_A^a(x) \partial J_B^a(w)\sim \frac{ 2 \dim g\; K_{AB}}{(x-w)^3}+(J_A^a \partial J_B^a)(w).
\end{equation}
We also used the identity
\begin{equation}
\frac{1}{2\pi i} \oint \frac{dx}{(x-w)^n} \frac{F(w)}{(z-x)^m}=\frac{(n+m-2)!}{(n-1)!(m-1)!} \frac{F(w)}{(z-w)^{n+m-1}}.
\end{equation}

Using
\begin{equation}
\sum_{A,B=1}^N G^{AB}K_{AB}=\sum_{A,B=1}^N\frac{K^{AB} K_{AB}}{2(1+hK^{AA})} = \sum_{A}^N  \frac{1}{2(1+hK^{AA})}.
\end{equation}
Then (\ref{TTope}) can be written as
\begin{equation}
T(z)T(w)\sim \frac{C(K)/2}{(z-w)^4}+\frac{2 T(w)}{(z-w)^2}+\frac{\partial T(w)}{(z-w)}
\end{equation}
where the central charge $C(K)$ is
\begin{equation}\label{centeralcharge}
C(K)=\sum_{A=1}^N\frac{\dim g}{1+hK^{AA}}.
\end{equation}
For $N=1$ and $K=k^{-1}$, we recover the usual value of $C$
\begin{equation}
C(k)=\frac{k\dim g}{h+k}.
\end{equation}
For the diagonal case $K_{AB}=k_A\delta_{AB}$, the central charge factorizes as expected
\begin{equation}
C(K)=\sum_{A=1}^N c\big(k_A\big)=\sum_{A=1}^N\frac{k_A\dim g}{h+k_A}.
\end{equation}

The Abelian limit of $\hat{g}_{N,K}$ corresponds to the algebra $\hat{u}_{N,K}=\hat{u}(1)^{\oplus_N}_K$, where $K$ is an $N\times N$ matrix-valued level. This algebra was studied in \cite{Gannon:1996hp} and its modular invariant partition functions were given.
In this case, the operator product expansion of the currents takes the form
\begin{equation}
J_A(z)J_B(w)\sim \frac{K_{AB}}{(z-w)^2}.
\end{equation}
This algebra was related to strings on complex multiplication tori in \cite{Nassar:2012dj}. In \cite{Gannon:1994sp}, Gannon studied the  classification of the modular invariant partition functions of the algebra $\widehat{su(2)}_{k_1}\oplus \cdots \oplus \widehat{su(2)}_{k_N} $ which in our notation corresponds to the algebra $\widehat{su(2)}_{N,K}$ with a diagonal $K$ matrix $K_{AB}=k_A \delta_{AB}$.

Now we go back to the algebra (\ref{matrix-level-alg}) and study it in more details.  By expanding the currents $J_A^a(z)$ in modes
\begin{equation}
J_A^a(z)=\sum_{m\in\mathbb{Z}}\frac{J_A^a(m)}{z^{m+1}}.
\end{equation}
The requirement  $J_A^a(z)^\dagger=J_A^a(z)$ leads to $J^a_A(m)^\dagger=J^a_A(-m)$.
We can write the OPE (\ref{matrix-level-alg}) in terms of the modes
\begin{equation}\label{moodalgebra}
\big[J_A^a(m),J_B^b(n)\big]=if_{abc}J_B^c(m+n)\delta_{AB}+mK_{AB}\delta_{ab}\delta_{m+n,0}.
\end{equation}

The central extension term $\Omega(J_A^a(m),J_B^b(n))=mK_{AB}\delta_{ab}\delta_{m+n,0}$ satisfies
\begin{equation}
\begin{gathered}
\Omega(J_A^a(m),J_B^b(n))=-\Omega(J_B^b(n),J_A^a(m)),\\[5pt]
\Omega\big(\big[J_A^a(m),\big[J_B^b(n),J_C^c(r)\big]\big]\big)+
\Omega\big(\big[J_C^c(r),\big[J_A^a(m),J_B^b(n)\big]\big]\big)\\[5pt]
+\Omega\big(\big[J_B^b(n),\big[J_C^c(r),J_A^a(m)\big]\big]\big)\\[5pt]
\qquad=if_{abc} \delta_{m+n+r,0}\big[m\delta_{BC} K_{AC}+n\delta_{AB} K_{CB}+r\delta_{CA} K_{AB}\big]
\end{gathered}
\end{equation}
This is equal zero if $K$ is a diagonal matrix and is non-zero for generic $K$. Hence, there is a cohomological obstruction for $\Omega(J_A^a(m),J_B^b(n))$ to define a $2$-cocycle on the loop group of $g^{\oplus N}$ and hence the algebra $\hat{g}_{N,K}$ is not the affine extension of the loop algebra of $g^{\oplus N}$.
 However, as pointed out in  \cite{Gannon:1996hp}, the important structure in CFT is the chiral algebra itself. The chiral algebra (\ref{matrix-level-alg}) is well defined and gives rise to unitary CFTs with  a set of genus-$1$ characters with definite modular transformation properties and a modular covariant partition function. The failure of the $2$-cocycle condition will not affect the OPE of the primary fields (defined by (\ref{prim-fields})) and as such  will not interfere with the fundamental
requirement of associativity of their correlation functions.


The modes of the energy-momentum tensor are given by
\begin{equation}
\mathbb{L}_n= \sum_{A,B=1}^N \sum_{m\in\mathbb{Z}} \frac{K^{AB}}{2(1+hK^{AA})} :J_A^a(m) J_B^a(n-m):.
\end{equation}
These modes satisfy the Virasoro algebra
\begin{equation}
\big[\mathbb{L}_n,\mathbb{L}_m\big]=(n-m) \mathbb{L}_{n+m}+\frac{C}{12}(n^3-n)\delta_{n+m,0},
\end{equation}
where the central charge $C$ is given by (\ref{centeralcharge}).

Primary fields are characterized by the following  OPE with the currents
\begin{equation}\label{prim-fields}
\begin{split}
J_A^a(z) \Phi_{\Lambda,\Xi}(w,\bar{w})& \sim \frac{-t_{\lambda_A}^a \Phi_{\Lambda,\Xi}(w,\bar{w})}{z-w}\\[4pt]
\bar{J}_A^a(\bar{z}) \Phi_{\Lambda,\Xi}(w,\bar{w}) &\sim \frac{\Phi_{\Lambda,\Xi}(w,\bar{w}) t_{\xi_A}^a }{z-w}
\end{split}
\end{equation}
This defines a primary field in the representation $(\Lambda,\Xi)$ of the horizontal subalgebra $\oplus_{A=1}^N g_A$ where
\begin{equation}
(\Lambda;\Xi)=(\oplus_A \lambda_A;\oplus_A \xi_A).
\end{equation}
A primary field $ \Phi_{\Lambda,\Xi}$  will give rise to a primary state $| \Phi_{\Lambda,\Xi}\rangle$ and the action of the current modes on primary states is
\begin{equation}
\begin{split}
J_A^a(0) | \Phi_{\Lambda,\Xi}\rangle &= -t_{\lambda_A}^a | \Phi_{\Lambda,\Xi}\rangle,\\
J_A^a(n) | \Phi_{\Lambda,\Xi}\rangle &=0,\quad n>0.
\end{split}
\end{equation}
These  states are also Virasoro primary states
\begin{equation}
\begin{split}
\mathbb{L}_0 | \Phi_{\Lambda,\Xi} \rangle &=h_\Lambda | \Phi_{\Lambda,\Xi} \rangle,\\[4pt]
\mathbb{L}_n | \Phi_{\Lambda,\Xi}\rangle &= 0, \quad n>0.
\end{split}
\end{equation}
where $h_\Lambda$ is the holomorphic conformal dimension of the primary field $\Phi_{\Lambda,\Xi}(w,\bar{w})$ and can be computed by noting that the action of $\mathbb{L}_0$ on primary states takes the simple form
\begin{equation}
\mathbb{L}_0= \sum_{A,B=1}^N \frac{K^{AB}}{2(1+hK^{AA})}:J_A^a(0) J_B^a(0):.
\end{equation}
This leads to
\begin{equation}
\begin{split}
\mathbb{L}_0  | \Phi_{\Lambda,\Xi} \rangle&=\sum_{A,B=1}^N \frac{K^{AB}}{2(1+hK^{AA})} t_{\lambda_A}^a t_{\lambda_B}^a  | \Phi_{\Lambda,\Xi} \rangle \\[4pt]
&=\bigg(\sum_{A=1}^N \frac{K^{AA}}{2(1+hK^{AA})} t_{\lambda_A}^a t_{\lambda_A}^a+\sum_{\substack{A,B=1 \\ A\neq B }}^N\frac{K^{AB}}{2(1+hK^{AA})} t_{\lambda_A}^a t_{\lambda_B}^a \bigg)  | \Phi_{\Lambda,\Xi} \rangle
\\[4pt]
&=\bigg( \sum_{A=1}^N \frac{K^{AA}}{2(1+hK^{AA})} C(\lambda_A)+\sum_{\substack{A,B=1 \\ A\neq B }}^N
\frac{K^{AB}}{2(1+hK^{AA})} H_{\lambda_A}^i H_{\lambda_B}^i \bigg)  | \Phi_{\Lambda,\Xi} \rangle\\[4pt]
&=\bigg( \sum_{A=1}^N \frac{K^{AA}}{2(1+hK^{AA})} C(\lambda_A)+\sum_{\substack{A,B=1 \\ A\neq B }}^N
\frac{K^{AB}}{2(1+hK^{AA})} \lambda_A \cdot \lambda_B \bigg)  | \Phi_{\Lambda,\Xi} \rangle
\end{split}
\end{equation}
where $C(\lambda_A)$ is the quadratic Casimir eigenvalue in the representation $\lambda_A$  and $H_{\lambda_A}^i$ are the elements of the Cartan subalgebra of $g_A$ in the representation $\lambda_A$ and we defined
\begin{equation}
\lambda_A \cdot \lambda_B = \lambda_A^i  \lambda_B^i.
\end{equation}
Hence the holomorphic conformal dimension of the field $\Phi_{\Lambda,\Xi}(w,\bar{w})$ is given by
\begin{equation}
h_\Lambda=\sum_{A=1}^N \frac{K^{AA}}{2(1+hK^{AA})} C(\lambda_A)+\sum_{\substack{A,B=1 \\ A\neq B }}^N
\frac{K^{AB}}{2(1+hK^{AA})} \lambda_A \cdot \lambda_B .
\end{equation}



Now we use the requirement of unitarity to derive some constraints on the matrix-level $K_{AB}$ and on the set of allowed representations. By restricting the algebra (\ref{moodalgebra}) to a subset of the generators one gets
\begin{equation}\label{KAA}
\big[J_A^a(m),J_A^a(-m)\big]=mK_{AA}.
\end{equation}
Using the fact that $J^a_A(m)^\dagger=J^a_A(-m)$ to define the state
\begin{equation}
|\Psi_A^a(m)\rangle=J_A^b(-m)|0\rangle, \quad m>0,
\end{equation}
where
\begin{equation}
J_A^b(m)|0\rangle=0, \quad m>0.
\end{equation}
Then (\ref{KAA}) can be written as
\begin{equation}
\langle\Psi_A^a(m) |\Psi_A^a(m)\rangle=mK_{AA}.
\end{equation}
Hence, unitarity forces us to chose
\begin{equation}\label{unitarity}
K_{AA}>0.
\end{equation}
By using the  Cartan-Weyl basis for the algebra (\ref{moodalgebra}) and in particular
\begin{equation}
[E^\alpha_A(1),E^{-\alpha}_A(-1)]=\alpha\cdot H_A(0)+K_{AA}
\end{equation}
One can derive the usual WZW bound
\begin{equation} \label{unitarity2}
\lambda_A\cdot \theta \leq K_{AA},
\end{equation}
where $\theta$ is the longest root of $g$.

Hence the positivity and integrality conditions on the diagonal elements $K_{AA}$ are required for the unitarity of the corresponding CFTs and in order to have a finite number of highest weight states as explained by (\ref{unitarity}) and (\ref{unitarity2}), respectively. It would be interesting to study extensions of these models for general $K_{AB}$ subject only to the ondition $K_{AA}\in \mathbb{Z}_+$.

Also, in a unitary CFT, the central charge and the conformal dimensions can't take negative values. This will lead to further constraints on $K^{AB}$ and can be analyzed in different cases.

\section{Genus-$1$ characters}\label{sec3}

The total Hilber space $\mathcal{H}$ of our CFT will split into subspaces $\mathcal{H}_{\Lambda}\otimes \mathcal{H}_{\bar{\Xi}}$
 \begin{equation}
 \mathcal{H}=M_{{\Lambda}\bar{\Xi}} \mathcal{H}_{\Lambda}\otimes \mathcal{H}_{\bar{\Xi}},
 \end{equation}
 where $M_{{\Lambda}\bar{\Xi}}\in \mathbb{Z}_+$ counts the multiplicity of the $\mathcal{H}_{\Lambda}\otimes \mathcal{H}_{\bar{\Xi}}$ in the CFT spectrum.
 On top of each one of the subspaces $\mathcal{H}_{\Lambda}\otimes \mathcal{H}_{\bar{\Xi}}$ sets a primary state $|\Lambda,\bar{\Xi}\rangle$. The other members of $\mathcal{H}_{\Lambda}$ will be generated by acting on $|\Lambda\rangle$ with the lowering operators of $\hat{g}_{N,K}$. The  level-$M$ descendent state is given by\footnote{Here we only consider  the holomorphic part of the CFT.}
\begin{equation}
|\Lambda,M\rangle=\prod_i J_{A_i}^{a_i}(-m_i) |\Lambda\rangle,\quad \sum_i m_i =M.
\end{equation}
Using the fact that
\begin{equation}
\big[\mathbb{L}_0,J_A^a(m)\big]=-m J_A^a(m)
\end{equation}
one can compute the conformal dimensions of $|\Lambda,M\rangle$
\begin{equation}\label{conformal-dimensions}
\mathbb{L}_0|\Lambda,M\rangle=(h_\Lambda+M)|\Lambda,M\rangle.
\end{equation}

To simplify the computation of the genus-$1$  characters  we write
\begin{equation}
\mathbb{L}_0=L_0+\tilde{L}_0,
\end{equation}
where
\begin{equation}
 \begin{split}
 L_0&=\sum_{A=1}^N\frac{K^{AA}}{2(1+hK^{AA})}:J_A^a(0) J_A^a(0):,\\[4pt]
  \tilde{L}_0& =\sum_{\substack{A,B=1 \\ A\neq B }}^N\frac{K^{AB}}{2(1+hK^{AA})}:J_A^a(0) J_B^a(0):.
 \end{split}
\end{equation}
These operators have the following action on primary states
\begin{equation}
 L_0 |\Lambda\rangle = h_\Lambda^0|\Lambda\rangle,\qquad   \tilde{L}_0 |\Lambda\rangle  = \tilde{h}_\Lambda |\Lambda\rangle,\qquad h_\Lambda = h_\Lambda^0+\tilde{h}_\Lambda,
\end{equation}
where
\begin{equation}\label{modularanomaly}
h_\Lambda^0=\sum_{A=1}^N \frac{K^{AA}}{2(1+hK^{AA})} C(\lambda_A),
\qquad
\tilde{h}_\Lambda=\sum_{\substack{A,B=1 \\ A\neq B }}^N
\frac{K^{AB}}{2(1+hK^{AA})} \lambda_A \cdot \lambda_B
\end{equation}

Using the the fact that
\begin{equation}
[L_0,\tilde{L}_0]=0.
\end{equation}
Hence
\begin{equation}
[\mathbb{L}_0,L_0]=0.
\end{equation}
This equation simply means that $\mathbb{L}_0$ and $L_0$ can be diagonalized simultaneously although with different eigenvalues. This observation simplifies the construction of the genus-one characters
\begin{equation}\label{characters}
\tilde{\chi}_\Lambda =\text{Tr}_{\mathcal{H}_\Lambda} q^{\mathbb{L}_0-c/24}.
\end{equation}
{{Using the action of $\mathbb{L}_0$ on the states $|\Lambda,M \rangle \in \mathcal{H}_\Lambda$  given by (\ref{conformal-dimensions}), the genus-one characters can be witten as
\begin{equation}
\begin{split}
\tilde{\chi}_\Lambda  & = \sum_M q^{h_\Lambda+M -c/24 }\\[2pt]
& = \sum_M q^{h_\Lambda^0+\tilde{h}_\Lambda+M -c/24 }\\[2pt]
& = q^{\tilde{h}_\Lambda} \sum_M q^{h_\Lambda^0+M -c/24 }\\[2pt]
& = q^{\tilde{h}_\Lambda} \chi_\Lambda
\end{split}
\end{equation}
where 
\begin{equation}
 \chi_\Lambda=\sum_M q^{h_\Lambda^0+M -c/24 }=\text{Tr}_{\mathcal{H}_\Lambda} q^{L_0-c/24} 
\end{equation}}}
is the affine characters of the diagonal matrix-level algebra $\hat{g}_{k_1}\oplus\cdots\oplus \hat{g}_{k_N}$ and is given by
\begin{equation}
\chi_\Lambda=\chi_{\lambda_1}^{k_1}\times\cdots \times\chi_{\lambda_N}^{k_N}.
\end{equation}
These characters have the following modular transformations\footnote{Here we only consider the specialized characters.}
\begin{equation}
\chi_\Lambda(-1/\tau)=\mathbb{S}_{\Lambda\Sigma} \; \chi_\Sigma (\tau),\qquad  \chi_\Lambda(-1/\tau)=\mathbb{T}_{\Lambda\Sigma} \; \chi_\Sigma (\tau),
\end{equation}
where
\begin{equation}
\mathbb{S}_{\Lambda\Sigma}=S^{k_1}_{\lambda_1\sigma_1}\otimes\cdots\otimes  S^{k_N}_{\lambda_N\sigma_N},\qquad
\mathbb{T}_{\Lambda\Sigma}=T^{k_1}_{\lambda_1\sigma_1}\otimes\cdots\otimes  T^{k_N}_{\lambda_N\sigma_N},
\end{equation}
where $S^{k_1}_{\lambda_1\sigma_1}$ and $T^{k_1}_{\lambda_1\sigma_1}$ are the modular matrices of the algebra $\hat{g}_{k_1}$.

Looking back at (\ref{characters}) and since $\tilde{\chi}_\Lambda=q^{\tilde{h}_\Lambda} \chi_\Lambda$, one learns that the matrix-level characters $\tilde{\chi}_\Lambda $ fulfils a representation of the modular group $SL(2,\mathbb{Z})$ up to a \textit{modular anomaly} given by $\tilde{h}_\Lambda$. It is interesting to note that the modular anomaly $\tilde{h}_\Lambda$ vanishes for a diagonal matrix level $K$, i.e., when there is no cohomological obstruction for $\hat{g}_{N,K}$ to define an affine extension of the loop algebra  of $g^{\oplus N}$.

\section{The algebra $\widehat{su(2)}_{N,K}$}\label{sec4}

To illustrate the results of the previous sections, we study the simplest kind of matrix-level algebras  which is based on $su(2)\oplus \cdots \oplus su(2)$. To
 derive a unitarity bound on the set of allowed representations we define the state
\begin{equation}
| \Theta_\Lambda \rangle=   J^{-}_A(-1)|\Lambda\rangle
\end{equation}
The norm of this state can be  computed using the mode algebra (\ref{moodalgebra})
\begin{equation}
\langle\Theta_\Lambda| \Theta_\Lambda \rangle= \langle \Lambda| J^{+}_A(1)J^{-}_A(-1)|\Lambda\rangle=\big(K_{AA}-\lambda_A\big).
\end{equation}
Hence the set of allowed representations should satisfy
\begin{equation}
\lambda_A \leq K_{AA}.
\end{equation}

The spectrum of conformal dimensions of the matrix-level algebra $\widehat{su(2)}_{N,K}$ is given by
\begin{equation}
h_\Lambda=\sum_{A=1}^N \frac{K^{AA}}{(1+hK^{AA}) } \frac{\lambda_A}{2}\bigg(\frac{\lambda_A}{2}+1\bigg) +\sum_{\substack{A,B=1 \\ A\neq B }}^N
\frac{K^{AB}}{2(1+hK^{AA})} \bigg(\frac{\lambda_A \lambda_B}{4}\bigg),
\end{equation}
where $\lambda_A$ is the Dynkin label of the highest weight state. Written in terms of the spin $j_A=\lambda_A/2$ of the highest weight state, the above formula can be written as
\begin{equation}
h_\Lambda=\sum_{A=1}^N \frac{K^{AA}}{(1+hK^{AA}) } j_A\big(j_A+1\big) +\sum_{\substack{A,B=1 \\ A\neq B }}^N
\frac{K^{AB}}{2(1+hK^{AA})} \big(j_A j_B\big).
\end{equation}

For $\widehat{su(2)}_{2,K}$ with the matrix level
\begin{equation}
K=
\begin{pmatrix}
1&1\\
1&2
\end{pmatrix}.
\end{equation}
The central charge is
\begin{equation}
C(K)=\frac{8}{5}.
\end{equation}
Notice that this central charge is not the same as the one for the algebra $\widehat{su(2)}_{1} \oplus \widehat{su(2)}_{2}$ (corresponding to a diagonal matrix-level $\text{diag}(1,2)$) which is equal to $5/2$.

The set of allowed representations of $\widehat{su(2)}_{2,K}$ is given by
\begin{equation}
P_K=\{(0;0),(0;1),(1;0),(1;1),(0;2),(1;2)\}
\end{equation}
The spectrum of holomorphic conformal dimensions is given by
\begin{equation}
h_{(0;0)}=0,\  h_{(0;1)}=\frac{1}{4},\  h_{(1;0)}=\frac{3}{10},\  h_{(1;1)}=\frac{17}{60},\  h_{(0;2)}=\frac{2}{3},\  h_{(1;2)}=\frac{13}{30}.
\end{equation}
Hence, the algebra $\widehat{su(2)}_{2,K}$ also has different values of the holomorphic conformal dimensions.





\section{Conclusion}

We studied an affine extension of the algebra  $g^{\oplus_N}=g\oplus\cdots\oplus g$ by spin-1 currents $J_A^a(z)$ with an OPE characterized by a matrix-valued level $K_{AB}$. The Sugawara construction were used to construct an energy-momentum tensor based on $J_A^a(z)$. This leads to a new classe of 2D conformal field theories with new values of the central charge and conformal dimensions. This algebra gives the familiar WZW models as a special case when the matrix-valued level is diagonal.
This affine extension is not the central extension of the loop algebra of $g^{\oplus_N}$ due to a cohomological obstruction which vanishes for diagonal $K$ matrices. A set of genus-$1$ characters were constructed which fulfils a representation of the modular group up to a modular anomaly which vanishes for diagonal $K$.

\section{Acknowledgement}

I am grateful to Mark Walton for useful discussions. I thank the annonymus refree for the valuable comments. This research is supported by the CFP at Zewail city of Science and Technology and by the STDF project 13858.



\section*{References}

\bibliographystyle{plain}

\bibliography{refs}
\end{document}